\documentclass[pre,twocolumn,superscriptaddress,floatfix]{revtex4}
\usepackage{bm,graphicx,amsmath,times}
\usepackage[mathscr]{eucal}

\newcommand{\E}[1]{Eq.~(\ref{#1})}
\newcommand{\F}[1]{Fig.~\ref{#1}}
\newcommand{\FF}[1]{Figure~\ref{#1}}

\newcommand{\T}[1]{TABLE~\ref{#1}}
\newcommand{\A}[1]{Appendix~\ref{#1}}

\newcommand{\avg}[1]{\left\langle#1\right\rangle}
\newcommand{\bs}[1]{\boldsymbol{#1}}
\newcommand{\psil}{\psi_\mathsf L}
\newcommand{\vphi}{\varphi}
\newcommand{\tvphi}{\tilde\vphi}
\newcommand{\tbeta}{\tilde\beta}
\newcommand{\tnu}{\tilde\nu}
\newcommand{\tmun}{\tilde\mu_{\mathrm{neut}}}

\newcommand{\tlambda}{\tilde\lambda}
\newcommand{\ddphi}{\nabla^2\vphi}

\newcommand{\R}{\mathscr{R}}

\newcommand{\ii}{{\rm i}}
\newcommand{\dd}{{\rm d}}
\newcommand{\ee}{{\rm e}}

\begin{document}

\title{Energy-Enstrophy Stability of  $\bs\beta$-plane Kolmogorov Flow with Drag}

\author{Yue-Kin Tsang}


\author{William R. Young}
\affiliation{Scripps Institution of Oceanography,
University of California, San Diego, La Jolla, California, 92093 USA}

\date{\today}

\begin{abstract}

We develop a new  nonlinear stability method, the Energy-Enstrophy (EZ) method, that is specialized to two-dimensional hydrodynamics;  the method is applied to a   $\beta$-plane flow driven by a sinusoidal body force, and retarded  by  drag  with damping time-scale $\mu^{-1}$. The standard energy method (Fukuta and Murakami, J. Phys. Soc. Japan, {\bf 64},  1995, pp 3725) shows that the laminar solution is monotonically and globally stable in a certain portion of the $(\mu,\beta)$-parameter space.  The EZ method  proves nonlinear stability in a larger portion of the $(\mu,\beta)$-parameter space. And by penalizing high wavenumbers, the EZ method  identifies a most strongly amplifying disturbance that is more physically realistic than that delivered by the  energy method.  Linear  instability calculations are used to determine the region of the $(\mu,\beta)$-parameter space where the flow is unstable to infinitesimal perturbations. There is only a small gap between the linearly unstable region and the nonlinearly stable region, and full numerical solutions show only small transient amplification in that gap. 

\end{abstract}



\maketitle

\section{Introduction}

 Kolmogorov flow is the simplest example of a two-dimensional motion forced at a single spatial scale. This provides an opportunity to understand the implications of the dual conservation laws of energy and enstrophy, the first of which is Fj{\o}rtoft's observation that nonlinear interactions transfer energy simultaneously  up and down scale \cite{Fjortoft53,Merilees75}.

The signature of Kolmogorov flow is that motion is maintained against dissipation by a sinusoidal body force.  Kolmogorov viewed his problem as an idealized example of a forced-dissipative system in which it might be possible to understand the  sequence of bifurcations resulting from increasing the Reynolds number \cite{Arnold91}. The first steps in this program were an analysis of the viscous linear stability problem \cite{Meshalkin61, Green74}. The weakly nonlinear theory, pivoted about the critical Reynolds number $\sqrt{2}$,  was subsequently developed\cite{Nepo76,Sivashinsky85}.

Laboratory experiments  using either soap films \cite{Rivera00, Burgess99}, or shallow  layers \cite{Bondarenko79}, can be driven by  electromagnetic forcing, or by the motion of an enveloping gas,  so that the body force  approximates a sinusoid. Thus  Kolmogorov's problem is also  important as an experimentally accessible flow in which  aspects of two-dimensional hydrodynamics can be tested.  In the laboratory the main dissipative mechanism is  drag on the adjacent walls, rather than lateral viscosity \cite{Dolzhansky87,Thess92}. 

 In  the geophysical context, the instability of planetary waves on a $\beta$-plane \cite{Lorenz72, Gill74, Lee03}  is similar to Kolmogorov's problem in many respects.  Ekman friction, which is equivalent to sidewall drag in the laboratory, also has a strong effect on the stability of planetary waves \cite{Dolzhansky85}.

Another geophysical motivation for studying the Kolmogorov problem is the equilibration of baroclinic turbulence.
The most unstable mode of  baroclinic instability is an exponentially growing  sinusoidal flow, that is  an exact solution of the equations of motion. The amplitude of this mode is   limited by  a secondary instability,  resembling  Kolmogorov instability. By deflecting energy into the barotropic mode, and thence into zonal jets, this secondary instability  equilibrates baroclinic turbulence by direct cascade of the thermal mode to high wavenumbers \cite{Salmon80}.   Thus weakly nonlinear $\beta$-plane Kolmogorov flow has been studied as a  model of zonal jet formation  in the geophysical context \cite{Manfroi99}.

Our main concern in this investigation is nonlinear stability analysis of Kolmogorov flow. This avenue was opened by Fukuta and Murakami \cite
{Fukuta95} using the energy method \cite{Joseph76,Drazin04}.  The energy method provides a sufficient condition for nonlinear stability by finding the critical value of the dissipation ensuring that  the disturbance energy  decreases \textit{monotonically} to zero. Our interest in this question is whether the second two-dimensional conservation law, namely squared vorticity or enstrophy, might be used to improve the nonlinear stability results in \cite
{Fukuta95}. 

In Sec.~\ref{formulation} we formulate the Kolmogorov stability problem. 
In Sec.~\ref{linstab} we discuss  the linear stability of the flow,  focusing on the limit in which the drag is much stronger than viscosity.  In Sec.~\ref{nonlinstab}, we extend the energy-stability condition of Fukuta and Murakami \cite{Fukuta95} to the $\beta$-plane and thus obtain a sufficient condition for nonlinear stability. Comparing the results of Sec.~\ref{linstab} with those of Sec.~\ref{nonlinstab}, we see that there is a region of parameter space in which the flow is linearly stable, but the energy method fails to prove nonlinear stability. In Sec.~\ref{enststab}  we develop a new  nonlinear stability method, the Energy-Enstrophy (EZ) method, which is specialized to two-dimensional hydrodynamics.  The EZ method provides a sufficient condition for stability which is stronger than the energy method, and consequently the gap between the results of linear stability and the nonlinearly stable region of parameter space is narrowed, but not eliminated. Section~\ref{conclude} concludes the paper.

\section{Formulation of the stability problem \label{formulation}}

With non-dimensional variables, the  vorticity equation is
\begin{equation}
\nabla^2 \psi_t + J(\psi,\nabla^2 \psi) + \beta \psi_x = \nu \nabla^4\psi - \mu \nabla^2 \psi + 
\cos(x-x_f)\, .
\label{qg1}
\end{equation}
In \E{qg1} the 
incompressible velocity field is obtained from a stream function $\psi(\bs x,t)$ according 
to $(u,v) = (-\psi_y, \psi_x)$. The domain is a doubly periodic square
$2 \pi L \times 2 \pi L$, where $L$ is an integer. The relative vorticity is $\nabla^2 \psi$, where $\nabla^2 \equiv \partial_x^2 + \partial_y^2$ is the two-dimensional Laplacian, and
$J(a,b)\equiv a_xb_y- a_yb_x$ is the Jacobian.  $\beta$ is the 
gradient of the Coriolis parameter along $y$ and the two dissipative mechanisms, 
viscosity $\nu$ and drag $\mu$, are represented by the first and second terms on 
the right hand side of \E{qg1}.

The flow in \E{qg1} is forced by a sinusoidal body-force,  which is the signature of the Kolmogorov flow.  In dimensional variables, the Kolmogorov forcing is specified as  
\begin{equation}
\tau_f^{-2} \cos k_f (x-x_f) \, ;
\end{equation}
 there is a length scale $k_f^{-1}$ and a time scale $\tau_f$. To obtain the non-dimensional form in \E{qg1}, we have scaled using $\tau_f$ and $k_f$. Thus, if  $*$ denotes a dimensional quantity, then in \E{qg1}  the  non-dimensional control parameters are
$\beta \equiv \tau_f\beta_*/k_f$, $\mu \equiv \tau_f \mu_* $,
$\nu \equiv k_f^2 \tau_f \nu_*$ and $L \equiv k_f L_*$.
Also in \E{qg1}, the phase
$
x_f \equiv \tan^{-1} \left[{\beta}/{(\mu+\nu})\right] 
$
 is defined so that there is a steady laminar solution:
 \begin{equation}
\psil(x) = - a \cos  x \, ,
\label{basic}
\end{equation}
where the amplitude of the laminar flow is
\begin{equation}
a(\beta,\mu,\nu) \equiv \frac{1}{\sqrt{\beta^2 + (\mu+\nu)^2}} \, .
\label{adef}
\end{equation}

\subsection{Dynamics of the  disturbance}

The disturbance $\vphi(\bs x,t)$ to the laminar stream function is defined by
\begin{equation}
\psi(\bs x,t) = \psil(x) + \vphi(\bs x,t)\, ,
\label{vphi}
\end{equation}
and the  equation of motion of $\vphi$ is obtained by substituting \E{vphi} into \E{qg1}
\begin{align} 
\ddphi_t +  J(\varphi, \ddphi) + a \sin x \, \left( \ddphi + \varphi \right)_y + \beta \vphi_x  \nonumber \\
= \nu \nabla^4 \vphi -\mu \ddphi \, .
\label{linphi}
\end{align}

  The disturbance energy $E_\vphi$ and enstrophy $Z_{\vphi}$ are defined  as
\begin{equation}
E_\vphi \equiv \frac12\avg{|\nabla\vphi|^2}  \quad \text{and} \quad Z_{\vphi} \equiv \frac12\avg{(\nabla^2\vphi)^2} \, ,
\label{EZdef}
\end{equation}
where $\avg{\cdots}$ denotes spatial average over the whole domain.
Multiplying \E{linphi} by $\vphi$ and   averaging gives the disturbance energy equation:
\begin{equation}
 \frac{\dd E_\vphi}{\dd t} = a\avg{\vphi_x\vphi_y\cos x}-2\mu E_\vphi -  2\nu Z_{\vphi}\, .
\label{dedt}
\end{equation}
Multiplying by $\nabla^2 \varphi$ and averaging produces the disturbance enstrophy equation: 
\begin{equation}
 \frac{\dd Z_\vphi}{\dd t} = a \avg{\vphi_x\vphi_y\cos x} - 2\mu Z_\vphi - 2\nu P_{\vphi}\, ,
\label{dzdt}
\end{equation}
where  
$
P_\vphi \equiv \frac12\avg{|\nabla\ddphi|^2}
$
is the disturbance palinstrophy.

It is remarkable that the disturbance energy and disturbance enstrophy are generated at equal rates i.e., the same source, namely  $a \avg{\vphi_x\vphi_y\cos x}$, appears in \E{dedt} and  in  \E{dzdt}. 
Thus subtracting \E{dzdt} from \E{dedt} gives
\begin{equation}
\frac12 \frac{\dd}{\dd t} (E_{\vphi} - Z_{\vphi}) = - \mu(E_{\vphi} -Z_{\vphi}) - \nu Z_{\vphi} + \nu P_{\vphi}\, .
\label{ddeltadt}
\end{equation}
This cancellation is a general property of flows forced by a single Helmholtz eigenmode, and is the basis for recent constraints on the spectral distribution of energy and enstrophy in two-dimensional turbulence \cite{Constantin94, Tran02, Alexakis06}. If $\nu=0$ then solution of \E{ddeltadt} shows that the difference  $E_{\vphi}(t) -  Z_{\vphi}(t)$ decays exponentially to zero. This observation is the basis of the EZ method in section \ref{enststab}.

\subsection{Linear instability, global stability and monotonic global stability}

\begin{figure} 
\includegraphics[width=0.41\textwidth]{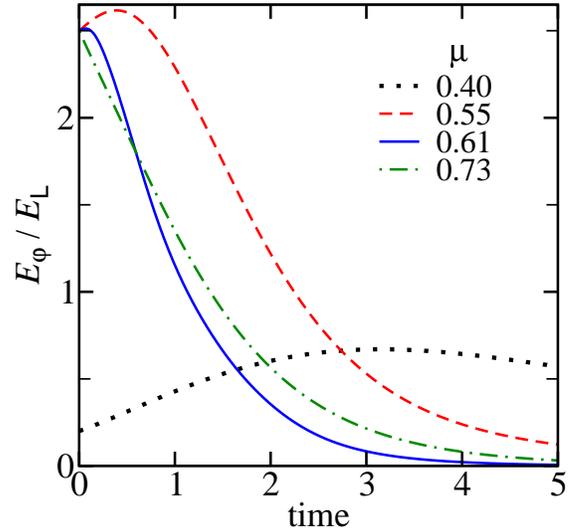}
\caption{(Color online) Disturbance energy $E_\vphi(t)$ from the numerical solution of \E{qg1} with $\beta=0$ and $\nu=10^{-3}$. The results are normalized by the laminar energy $E_{\mathsf L}=\frac14 a^2$. The solid and dashed curves illustrate the transient growth in $E_\vphi$ when the laminar solution $\psil(x)$ is globally stable, but not monotonically globally stable. The dot-dashed curve illustrates monotonic global stability, and the dotted curve shows the development of a linearly unstable perturbation.}
\label{asystab}
\end{figure} 

Following Joseph \cite{Joseph76}, we say that the laminar solution $\psil(x)$ is \textit{globally stable} if for all initial disturbances
\begin{equation}
\lim_{t \to \infty} E_{\varphi}(t) =0\, .
\end{equation}
 An even stronger form of stability is \textit{monotonic global stability}, meaning that  $\dd E_{\vphi}/\dd t \leq0$ for all $t \geq 0$. In both cases the laminar solution ultimately attracts all initial conditons. 

Global stability does not forbid transient increases in $E_{\vphi}(t)$ e.g., see \F{asystab}.
It is a limitation of the standard  energy method, reviewed  in  section \ref{nonlinstab}, that  only   monotonic global stability can be established.
 Using the EZ method, we show in section \ref{enststab} that there is a region of parameter space within which Kolmogorov flow  is globally stable, but not monotonically globally stable. The solid and dashed curves in \F{asystab} illustrate this situation.

A flow is said to be linearly unstable if the linearized version of \E{linphi} has exponentially growing eigensolutions. This is the subject of 
section \ref{linstab}.

\section{Linear instability}
\label{linstab}

Our goal in this section is to find the ``neutral surface" in the $(\mu,\beta,\nu)$-space below which $\psil(x)$ is linearly unstable.
  We linearize \E{linphi} and write $\vphi$ in Floquet form \cite{
Gill74,Thess92}:
\begin{equation}
\vphi(\bs x,t) = \Re\left\{ e^{\ii(kx + ly-\omega t)} \tvphi(x) \right\}\, ,
\label{floquet}
\end{equation}
where $-1/2<k\leq1/2$ and $\tvphi(x)$ has the same periodicity as $\psil(x)$, i.e., $\tvphi(x)=\tvphi(x+2\pi)$. The resulting eigenproblem is
\begin{align}
\left[\ii\tnu(D^2-l^2)^2+l\sin x(D^2-l^2+1) - \ii\tbeta D \right]\tvphi\nonumber \\
=\tlambda(D^2-l^2)\tvphi \, ,
\label{lineigen}
\end{align}
where the differential operator $D$ and the eigenvalue $\tlambda$ are
\begin{equation}
D \equiv \frac{\dd}{\dd x} + \ii k\, , \qquad \tlambda\equiv\frac{\omega+\ii\mu}{a}\, .
\label{dd}
\end{equation}
The two parameters in \E{lineigen} are defined as
\begin{equation}
\tbeta\equiv\frac{\beta}{a} \, , \qquad\tnu\equiv\frac{\nu}{a}\, .
\label{para}
\end{equation}
The basic flow is linearly unstable if there exists an $\omega$ with positive imaginary part, i.e. $\Im\{\omega\}=a\Im\{\tlambda\}-\mu>0$.

\subsection{Tracing the neutral surface}

\begin{figure*} 
\includegraphics[width=0.95\textwidth]{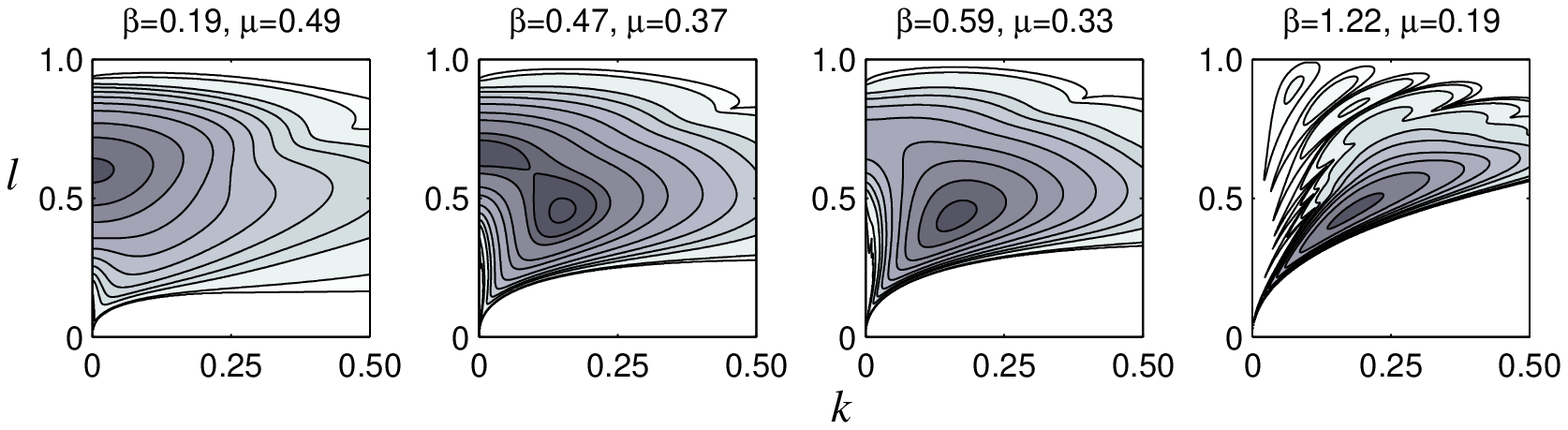}
\caption{(Color online) This shows $\max_n \Im\{\tlambda_n(k,l;\tbeta,\tnu=0)\}$ as a function of $k$ and $l$; dark areas are regions of large values. The values of $(\mu,\beta)$ in the titles are obtained by substituting $\tbeta$ and the peak value, $\tmun$, into \E{para1} and \E{para2}. }
\label{imageigen}
\end{figure*}

 For a given pair of $(\tbeta,\tnu)$, we solve \E{lineigen} numerically to obtain the eigenvalue spectrum $\{\tlambda_n(k,l;\tbeta,\tnu)\}$ as a function of the wavenumber $(k,l)$. The integer $n$ on $\tlambda_n$ indexes  the eigenbranch. Once we possess the spectrum, we define  the function
\begin{equation}
\tmun(\tbeta,\tnu) \equiv \max_{k,l,n} \Im\{\tlambda_n(k,l; \tbeta,\tnu)\}\, .
\label{mustar}
\end{equation}
\FF{imageigen} shows $\max_n \Im\{\tlambda_n(k,l;\tbeta,\tnu)\}$ at four values of $\mu$ and $\beta$ which happen to fall on the inviscid neutral curve. The function $\tmun(\tbeta,\tnu)$ in \E{mustar} is obtained by searching through the $(k,l)$-plane in Figure \ref{imageigen} to find the maximum. 

Because  $a\tmun-\mu$ is the growth rate of the most unstable mode,
the neutral surface in the $(\mu, \beta, \nu)$-space is defined by 
\begin{equation}
\mu=a\tmun(\tbeta,\tnu)\, .
\label{linneut}
\end{equation}
Given $\tmun(\tbeta,\tnu)$, the 
 neutral surface is traced using the parametric equations:
 \begin{subequations}
\begin{align}
\beta &= \frac{\tbeta}{[\tbeta^2+(\tmun+\tnu)^2]^{1/4}}\,, \label{para1} \\
\mu &= \frac{\tmun}{[\tbeta^2+(\tmun+\tnu)^2]^{1/4}}\,, \label{para2} \\
\nu &= \frac{\tnu}{[\tbeta^2+(\tmun+\tnu)^2]^{1/4}}\, . \label{para3}
\end{align}
\end{subequations}
The expressions above follow from \E{adef},  \E{para} and \E{linneut}.
The neutral surface is obtained from \E{para1} through \E{para3} by varying the parameters $(\tbeta,\tnu)$  between $0$ and $\infty$.

\begin{figure} 
\includegraphics[width=0.4\textwidth]{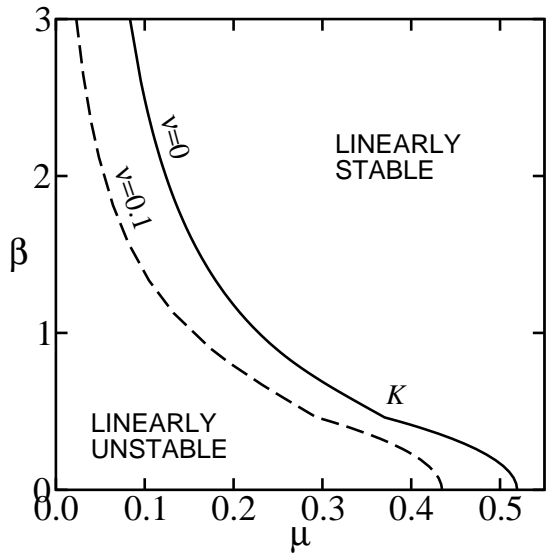}
\caption{Results from linear stability analysis of \E{qg1}. For the inviscid case $\nu=0$,
the solid line is the neutral curve. The point $K$ is where the most unstable mode changes character. For $\nu=0.1$, the dashed line is the neutral curve.}
\label{linstbcurve}
\end{figure}

\subsection{Gill's inequality}

In the inviscid case, $\nu=0$, an argument of  Gill \cite{Gill74} and Lorenz \cite{Lorenz72} shows that it is sufficient to search for unstable modes within the circle,
\begin{equation}
k^2+l^2<1\, .
\label{gl}
\end{equation}
 In \A{GillzAppen} we show that  \E{gl} also applies to linearly unstable modes of the viscous problem.   Gill's inequality underscores Fj{\o}rtoft's result \cite{Fjortoft53} by  showing that a growing eigenmode must transfer energy to spatial scales which are larger  than that of the laminar flow in \E{basic}.

\subsection{The inviscid case, $\nu=0$}

The  procedure outlined above is particularly simple in the inviscid case $\nu=0$. The inviscid neutral curve in the $(\mu,\beta)$-plane is the solid curve in  \F{linstbcurve}. The neutral curve intersects the $\mu$-axis at $\mu =0.52$.  After disentangling  notational differences,  $\mu =0.52$ agrees with Thess's value for unbounded Kolmogorov flow \cite{Thess92}. 

An interesting feature of the neutral curve is the appearance of a kink at the point $K$:$(\mu,\beta)\approx(0.37,0.46)$. This is due to a change in character of the most unstable mode,  evident between the first and third panels of  \F{imageigen}. 
As one moves along the neutral curve by decreasing $\mu$, the peak eigenvalue on the $k=0$ line is getting smaller and is eventually overtaken by another peak emerging at a location with $k\neq 0$. Consequently the wavenumber $k$ of the most unstable mode discontinuously  jumps from $k=0$ to   $k=0.15$ at $K$. \FF{mukl} shows the transition by plotting the variation of $\tmun$ and the wavenumber of the most unstable mode along the neutral curve.
\begin{figure} 
\includegraphics[width=0.42\textwidth]{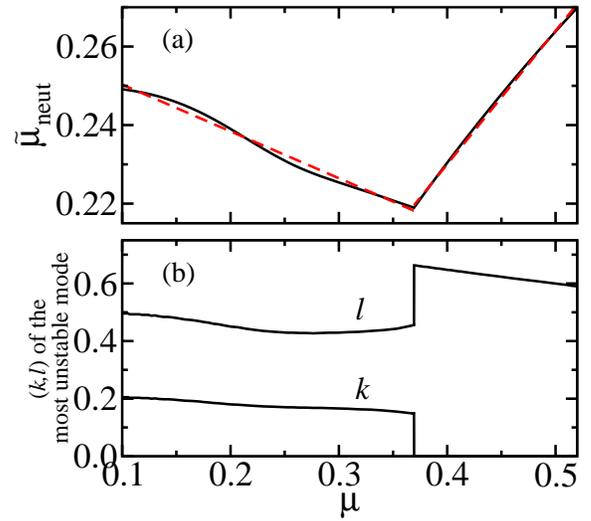}
\caption{(Color online) Variation of (a) $\tmun$, defined in \E{mustar}, and (b) wavenumber $(k,l)$ of the most unstable mode along the linear stability neutral curve. The dashed lines in (a) show the piecewise linear  approximation \E{tmuapp}.}
\label{mukl}
\end{figure}

We  find that for all growing modes, except for those with $k=0$, the real part of $\omega$ is non-zero, $\Re\{\omega\}\neq 0$. Hence the jump in $k$ to a non-zero value as $\mu$ decreases  also signals that the most unstable mode becomes a traveling wave. In summary, along the neutral curve in \F{linstbcurve}, the most unstable modes to the right of point $K$ are stationary disturbances with $k=0$, and those on the left of point $K$ are traveling waves with $k\neq 0$.

We  mention that
in the inviscid case, the neutral curve can be written as
\begin{equation}
\beta=\sqrt{\frac{\tmun^2}{\mu^2}-\mu^2}\, ,
\label{neutnu0}
\end{equation}
and \F{mukl}(a) indicates that
\begin{subequations}
\begin{align}
\tmun(\mu) &\approx 0.26-0.12\mu\, , &0&<\mu\leq0.37\,,\\
\tmun(\mu) &\approx 0.09+0.34\mu\, , &0.37&<\mu\leq0.52\,.
\end{align}
\label{tmuapp}%
\end{subequations}
Substituting \E{tmuapp} into \E{neutnu0} gives an approximate expression for the $\nu=0$ neutral curve; this approximation matches the numerical result to within the line width in \F{linstbcurve}.

\subsection{The effects of viscosity}

When $\nu\neq 0$, the neutral curve can be traced out by a similar procedure. The dashed curve in \F{linstbcurve} is the result for $\nu=0.1$, showing the stabilizing effect of $\nu$. For $\nu=10^{-3}$, the neutral curve is indisguishable from the inviscid curve, indicating the limit $\nu\to 0$ is non-singular. We find a kink $K$ in  the neutral curve, and a corresponding change in the structure of the most unstable mode,  always appears at $\beta\sim0.46$, independent of $\nu$.

\section{The Energy Method}
\label{nonlinstab}

We now turn to  the  stability of the basic flow \E{basic} using the energy method. Fukuta and Murakami \cite{Fukuta95} have previously considered this problem with $\beta=0$. It is  easy to adapt their results to nonzero $\beta$ because the disturbance energy equation does not involve $\beta$: $\beta$ enters the energy method only through the laminar amplitude, $a(\beta,\mu,\nu)$ in \E{adef}.

We begin by rewriting the disturbance energy equation \E{dedt} as
\begin{equation}
\frac{\dd E_\vphi}{\dd t} = 2 \left(a\R[\vphi]-\mu\right)E_\vphi -  2\nu Z_{\vphi}\, ,
\label{dedt1}
\end{equation}
where
\begin{equation}
\R[\vphi] \equiv \frac{\avg{\vphi_x\vphi_y\cos x}}{\avg{|\nabla\vphi|^2}}\, .
\label{danger}
\end{equation}
The homogeneous functional $\R$ represents the transfer of energy by Reynolds' stresses between the basic flow and the perturbation flow. 

The basic idea of the energy method is that if the dissipative parameters $\mu$ and $\nu$ are large enough, then the right hand side of \E{dedt1} is negative for all possible $\varphi$'s (not just $\varphi$'s which happen to satisfy the equations of motion). This implies that the disturbance energy decreases monotonically to zero, and that the laminar solution is monotonically globally stable.

For instance, since
\begin{equation}
\R[\varphi] \leq  \frac{\avg{|\vphi_x||\vphi_y|}}{\avg{|\nabla\vphi|^2}}\leq \frac{1}{2}\, ,
\label{simp1}
\end{equation}
it follows  from \E{dedt1} that 
\begin{equation}
\frac{\dd E_\vphi}{\dd t} \leq 2 \left( \frac{a}{2}  - \mu \right)E_\vphi \, ,
\label{simp2}
\end{equation}
and Gronwall's inequality implies that 
\begin{equation}
E_{\vphi}(t) \leq \ee^{(a-2 \mu)t}\, .
\end{equation}
That is, if $a < 2 \mu$ then the laminar flow in \E{basic} is monotonically globally stable. This conclusion relies on the seemingly  crude inequality in \E{simp1} and one expects that a stronger condition for monotonic global stability might be obtained by working harder and solving the variational problem suggested by maximizing the right hand side of \E{dedt1}. In this spirit we recapitulate some of the variational results in \cite{Fukuta95} by discussing the simplest case $\nu=0$ in some detail.

\subsection{The inviscid case, $\nu=0$}

With $\nu=0$, let $\R_E$ be the maximum of $\R[\vphi]$ over all function $\vphi(\bs x)$ satisfying the periodic boundary conditions:
\begin{equation}
\R_E \equiv \max_\vphi\R[\vphi]\, .
\label{dstar}
\end{equation}
According to the energy-stability method, the function $\vphi_E(\bs x)$ which maximizes $\R$ is the ``most dangerous disturbance'', meaning the most efficient energy-releasing disturbance.

Applying  Gronwall's inequality to \E{dedt1}, if $a\R_E < \mu$ then  $E_\vphi$ decays monotonically to zero and the laminar solution is monotonically globally stable. Hence with $\nu=0$, the energy-stability boundary in the $(\mu,\beta)$-plane is given by $\mu=a\R_E$, or
\begin{equation}
\beta=\sqrt{\frac{\R_E^2}{\mu^2}-\mu^2}\, .
\label{stabcurve1}
\end{equation}
Since $\R[\vphi]$ is a homogeneous functional, we can find $\R_E$ by maximizing$
 \avg{\vphi_x\vphi_y\cos x}
$
subject to the constraint $\avg{|\nabla\vphi|^2}=1$. This leads to the  Euler-Largrange equation,
\begin{equation}
(D^2-l^2)\tvphi  = \ii l\lambda \left( \cos x\,D- \frac12\sin x \right)\tvphi \, ,
\label{eigen1}
\end{equation}
where  $D$ is the differential operator defined in \E{dd} and the  Lagrange multiplier appears as a real eigenvalue $\lambda$.
 Then $\R_E$ is given by $1/\lambda_{\min}$, where $\lambda_{\min}$ is the minimum eigenvalue.

Again we represent $\varphi$ using  the Floquet form in \E{floquet} (but now with $\omega=0$) and  solve \E{eigen1} numerically to obtain $\lambda_0(k,l)$, the eigenvalue of the gravest mode, as a function of $k$ and $l$. \FF{eigen_lk_nu0} displays $\lambda_0(k,l)$
\begin{figure} 
\includegraphics[width=0.47\textwidth]{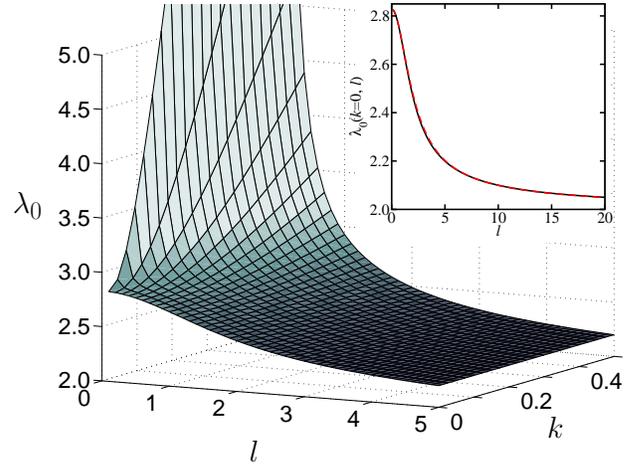}
\caption{(Color online) Eigenvalue of the gravest mode of \E{eigen1}, $\lambda_0(k,l)$. The minimum is at $k=0$ and $l\to\infty$. The inset shows a cut along $k=0$ (solid line); $\lambda_0(0,l)$ decreases monotonically from $2\sqrt{2}$ at $l=0$ to $2$ as $l\to\infty$. The dashed line is the  approximation \E{billzkluge}.}
\label{eigen_lk_nu0}
\end{figure}
above the $(k,l)$-plane and the conclusion is that
\begin{equation}
\lambda_{\min}=\min_{k,l}\lambda_0(k,l)=2\, ,
\label{lambmin}
\end{equation}
with the minimum  achieved at $k=0$ and $l\rightarrow\infty$. The numerical result \E{lambmin} is supported by the analysis of \E{eigen1} in \A{nu0}.  

The conclusion $\R_E=1/\lambda_{\min}=1/2$ is anticipated precisely by the simple  inequality in \E{simp1}: in the inviscid case the variational solution does not improve the energy stability boundary at all. The only reward from the variational solution is that it provides the form of the most dangerous  disturbance $\vphi_E(\bs x)$. The analysis in  \A{nu0} suggests that this disturbance can be approximated by the trial function
\begin{equation}
\vphi_E(\bs x) \approx \lim_{l \to \infty} \cos\left[ l (y + \sin x)  \right] \exp\left( \frac{l}{ 2}  \cos 2x \right)\,.
\label{mostdangerous}
\end{equation}
\F{E_dangerous} shows that $\R$ evaluated at the trial function above indeed approaches one-half as $l \to \infty$.
\begin{figure} 
\includegraphics[width=0.47\textwidth]{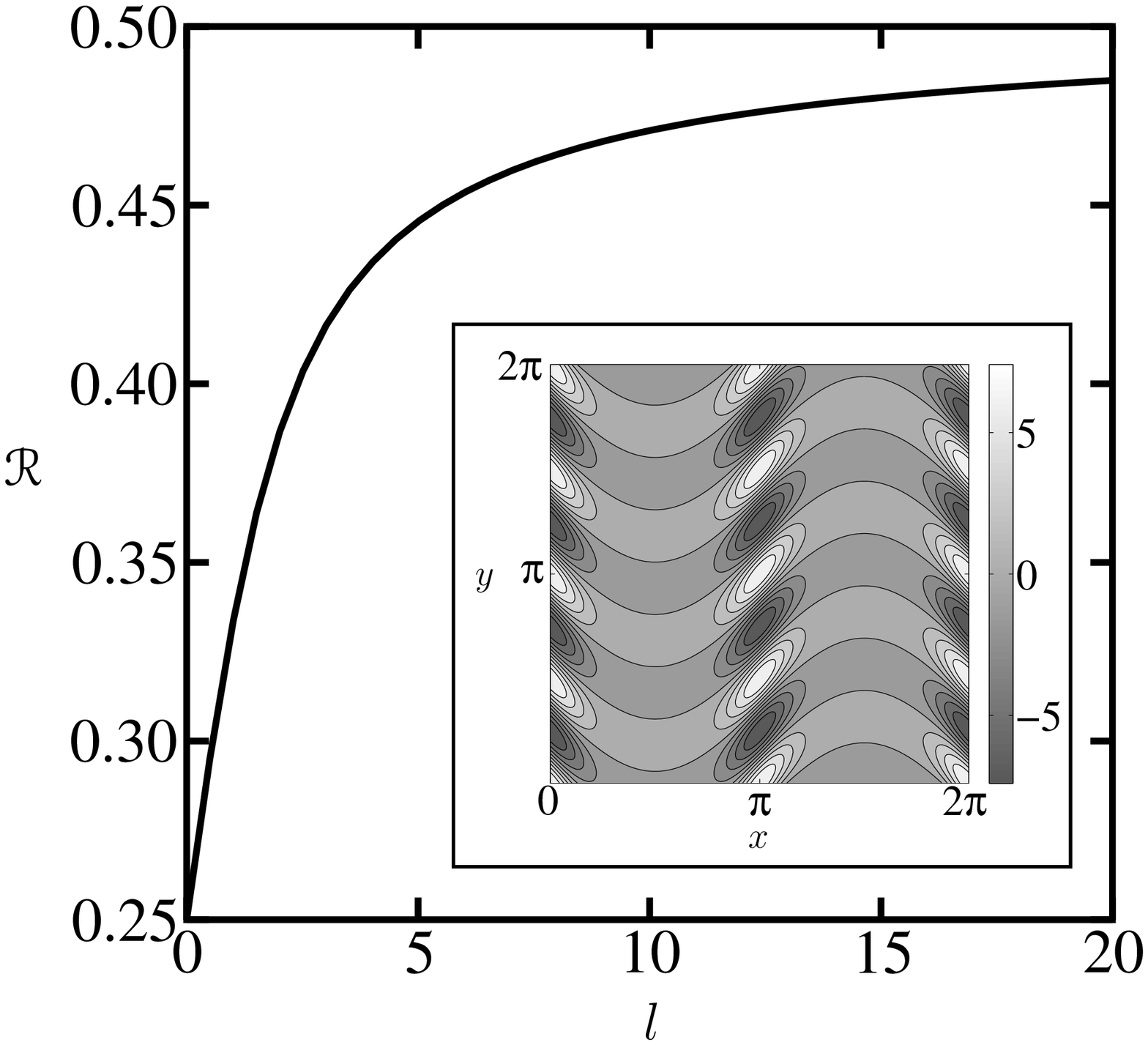}
\caption{The functional $\R$ (\E{danger}) evaluated at the trial function \E{mostdangerous} with different $l$. The inset shows the trial function with $l=4$.}
\label{E_dangerous}
\end{figure}

The Reynolds' stresses of the most dangerous disturbance are concentrated in the neighbourhood of $\cos x =\pm1$. Since there is no penalty attached to using very small spatial scales in the functional $\R[\vphi]$, this concentration can be made ever more intense by taking $l \to \infty$. The trial function shown in the inset of \F{E_dangerous} illustrates this strategy for maximizing $\R[\vphi]$. But $\vphi_E(\bs x)$ is realized  only in the $l\to \infty$ limit, and this is distinctly unphysical. This point is discussed further  in Section \ref{enststab}.

With $\R_E=1/2$, it follows from \E{stabcurve1} that the inviscid energy-stability boundary is given by
\begin{equation}
\beta=\sqrt{\frac{1}{4\mu^2}-\mu^2}\, .
\label{estab}
\end{equation}
\E{estab} is plotted as a dotted curve in \F{stabcurves}.
\begin{figure} 
\includegraphics[width=0.4\textwidth]{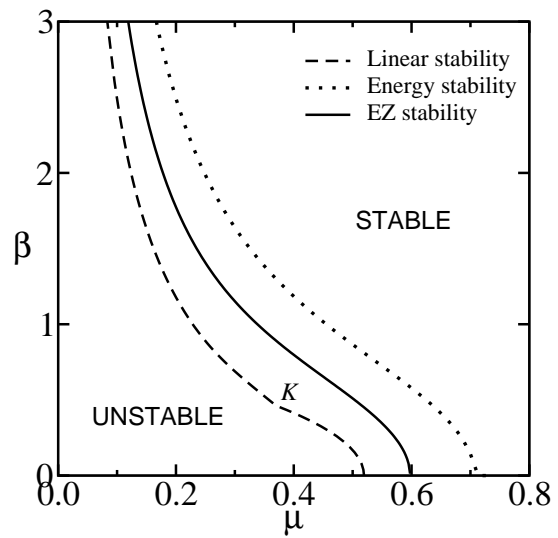}
\caption{A summary showing three different stability boundaries for two-dimensional Kolmogorov flow. The dashed curve is the neutral curve of the linear stability documented in Section \ref{linstab}; the dotted curve is the energy stability boundary in \E{estab}; the solid curve is the EZ stability boundary in \E{ezstabcurve}. The point $K$ at $(\mu,\beta) = (0.37,0.46)$ is where the most unstable linear mode changes from a stationary disturbance to a traveling wave.}
\label{stabcurves}
\end{figure}
Despite the pathology discussed above, it is remarkable that the energy-stability boundary parallels the linear-stability neutral curve of Section \ref{linstab} even though  the parameter $\beta$ enters the energy method  only through the function $a(\beta,\mu,\nu)$ in \E{adef}. This is an indication that the grossest features of the instability are determined by the amplitude of the laminar solution and the disturbance energy equation \E{dedt1}.

\subsection{The viscous case}

\begin{figure} 
\includegraphics[width=0.42\textwidth]{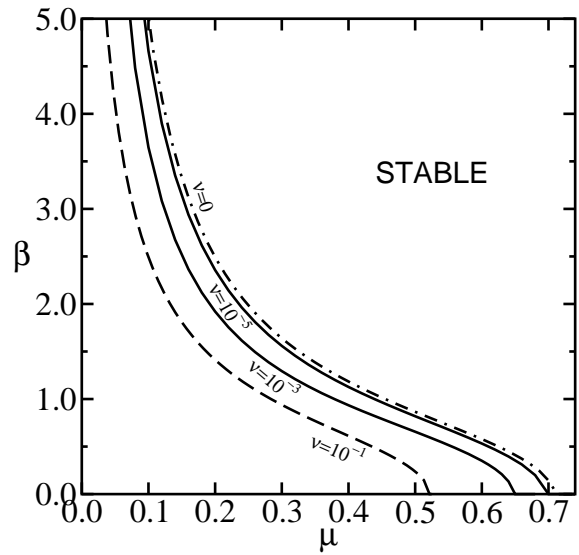}
\caption{This shows energy-stability boundaries in the $(\mu,\beta)$ parameter plane. The flow is monotonically globally stable if the parameters $(\beta,\mu,\nu)$ locate it above the appropriate curve.  }
\label{curvy}
\end{figure}

\begin{figure} 
\includegraphics[width=0.42\textwidth]{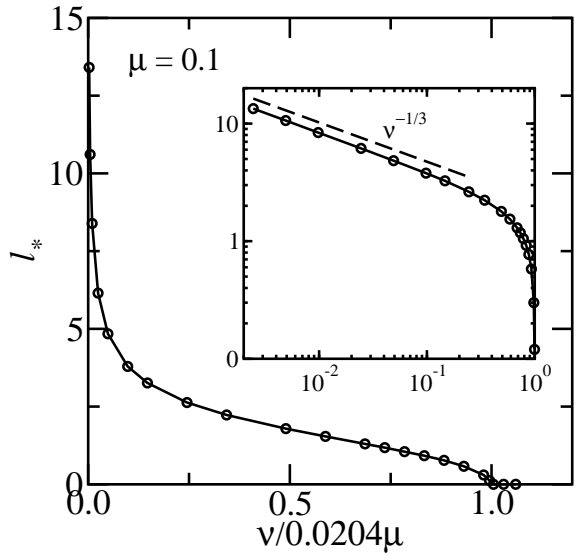}
\caption{This  shows the dependence of the wavenumber $l_*$ of the most dangerous disturbance on $\nu$. (The most dangerous disturbance has $k_*=0$ for all $\nu$.) If $\nu\geq 0.0204 \mu$ then $l_*=0$. As $\nu$ decreases, $l_*$ increases and ultimately, as show in the insert,  $l_* \sim \nu^{-1/3}$.}
\label{lvsnu}
\end{figure}

Turning now to $\nu \neq 0$, the main point of interest is how small viscosity affects the energy-stability curve in \F{stabcurves}. In this case the solution of the variational problem can provide a substantial improvement over the simple inequality  \E{simp1}. Following Fukuta and Murakami \cite{Fukuta95}, the viscous version of \E{eigen1} can be solved numerically and the results are shown in \F{curvy}. As $\nu \to 0$ the energy-stability curves limit to the inviscid curve in \E{estab}. Thus the limit $\nu \to 0$ is not singular.

One other point of interest is the wavenumber $l_*$ of the most dangerous disturbance.    \F{lvsnu} shows the dependence of $l_*$ on $\nu$.  In agreement with the small-wavenumber expansion  \cite{Fukuta95}, if $\nu/\mu >0.0204$ then $l_*=0$ and $\R_E=1/\sqrt{8(1+\nu/\mu)}$. Additional asymptotic analysis of ours (not presented here)  shows that if $\nu/\mu \ll 0.0204$ then
\begin{equation}
l_* = \left(\frac{\mu}{2 \nu} \right)^{1/3} + O\left(\nu^0\right)\, .
\end{equation}
In the limit $\nu \to 0$ we recover the inviscid result $l_*= \infty$.

\section{The Energy-Enstrophy Method \label{enststab}}

The gap between the neutral curve and the energy-stability boundary in \F{stabcurves} reflects the stark difference between the linearly unstable eigenmodes and the most efficient energy-releasing disturbance identified by the energy method. Specifically, the wavenumbers of the exponentially growing linear modes satisfy  Gill's inequality \E{gl} while, according to the energy method, the most efficient energy-releasing disturbance has a wavenumber $(k,l)=(0,\infty)$.  To improve on the energy method, the energy-enstrophy (EZ) method  uses both  the disturbance energy and  enstrophy.
 
 \subsection{The EZ method, $\nu=0$}

With $\nu=0$, \E{ddeltadt} implies that
\begin{equation}
E_\vphi(t) -Z_\vphi(t) =  e^{-2\mu t}\left[ E_\vphi(0) -Z_\vphi(0)\right]\, .
\label{attractor}
\end{equation}
Thus, although an arbitrary initial condition may not fall in the set
\begin{equation}
EZ \equiv \Big \lbrace \vphi(\bs x):\, E_{\vphi}=Z_{\vphi} \Big\rbrace\, ,
\label{EZset}
\end{equation}
the solution to \E{linphi} (with $\nu$=0) is attracted by $EZ$. And if the initial disturbance does happen to fall within $EZ$ then it stays within $EZ$. We refer to initial conditions which fall in $EZ$ as   ``$EZ$-disturbances''. 

\FF{eztraj} illustrates the approach to the attracting set $EZ$ using a numerical solution of \E{linphi}. Panel (a) shows that solutions starting outside of $EZ$ are quickly attracted to  $EZ$; panels (b) and (c) show that solutions starting in $EZ$, remain close to $EZ$. The trajectories with $\mu=0.4$ in panels (a) and (b) show two unstable solutions; nonlinearity halts the growth of the disturbance so that both trajectories asymptote to $(E_{\vphi} , Z_{\vphi} )\approx (0.49, 0.49)$ as $t \to \infty$. The solutions with $\mu=0.61$ in panels (a) and (c) show disturbances whose energy decays monotonically to zero (although $\mu=0.61$ is below the energy stability boundary).  These numerical results, with $0< \nu \ll \mu$, show that $EZ$ is an attracting set with small but nonzero viscosity.

To obtain an energy-stability bound for $EZ$-disturbances we seek
\begin{equation}
\R_{EZ} \equiv \max_{\vphi\in EZ}\R[\vphi]\, .
\label{dstar1}
\end{equation}
This is equivalent to  maximizing $\R[\vphi]$  in \E{danger} subject to the constraints
\begin{equation}
 \avg{|\nabla\vphi|^2} = \avg{(\ddphi)^2} = 1 \, .
\label{twoconstraints}
\end{equation}
The solution $\vphi_{EZ}(\bs x)$ is the most dangerous $EZ$-disturbance.

The search in \E{dstar1} is over a smaller set than the one used to define $\R_E$ in \E{dstar}, and so $\R_{EZ} \leq \R_E$. In anticipation of the variational calculation in the next subsection, we remark that
\begin{equation}
\R_{EZ} = 0.3571 < \R_E = \frac12\, .
\label{Rstarstar}
\end{equation}
Furthermore, the argument surrounding \E{gill3} shows that the Floquet wavenumber of an $EZ$-disturbance  must satisfy Gill's inequality \E{gl}. Thus the pathology of the energy method is avoided.

The  disturbance energy of an $EZ$-disturbance satisfies 
\begin{equation}
\frac{\dd E_{\varphi}}{\dd t} \leq 2(a \R_{EZ} - \mu) E_{\vphi}\, ,
\end{equation}
and Gronwall assures us that the energy decays monotonically to zero provided that $a \R_{EZ} < \mu$. Using the definition of $a$ in \E{adef}, this condition leads to the EZ stability curve 
\begin{equation}
\beta = \sqrt{ \frac{ \R^2_{EZ} }{\mu^2} - \mu^2}\, ,
\label{ezstabcurve}
\end{equation}
shown as a solid curve in \F{stabcurves}.  The trajectories with $\mu=0.61$ in panels (a) and (c) of \F{eztraj} show two solutions just to the right of the EZ stability curve in \F{stabcurves}.

\begin{figure} 
\includegraphics[width=0.42\textwidth]{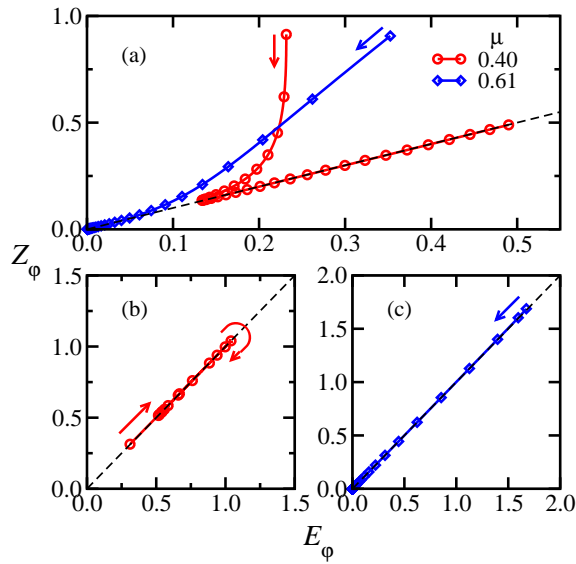}
\caption{(Color online) Trajectories of some illustrative solutions of \E{linphi} in the $[E_{\vphi}(t) , Z_{\vphi}(t)]$-plane; all solutions have $\beta=0$ and $10^{-3} \leq \nu \leq 5\times 10^{-3}$. The dashed line  marks the set $EZ$ in \E{EZset}. }
\label{eztraj}
\end{figure}

In \F{stabcurves} there is a  gap between the EZ curve and the energy-stability curve. In this region  $EZ$-disturbances decay monotonically, while general initial disturbances might have amplifying disturbance energy. In fact, initial amplification can be assured by deploying the trial function on the right of  \E{mostdangerous} at $t=0$. We argue heuristically that this amplification can only be transient: \E{attractor} shows that the set in \E{EZset} attracts all initial conditions. Thus, since $EZ$-disturbances decay monotonically to zero, general initial disturbances will also eventually decay to zero as they evolve towards $EZ$ via \E{attractor}.  We are arguing heuristically that in the gap between the EZ curve and the energy-stability curve the laminar solution is globally stable, but not monotonically globally stable.

\subsection{The EZ variational problem}
\label{ezvari}

To obtain $\R_{EZ}=0.3571$ in \E{Rstarstar}, we maximize $\avg{\vphi_x\vphi_y \cos x}$ subject to the constraints that the disturbance enstrophy and energy are both equal to one-half. A direct assault based on solving the Euler-Lagrange equation is described in \A{EZappen}. It is more instructive to develop some intuition using the  trial function
\begin{equation}
\vphi = A_0 \cos ly - B_1 \sin ly \sin x\, ,
\label{trial1}
\end{equation}
for which
\begin{equation}
\avg {\vphi_x \vphi_y \cos x} = \frac14 l A_0B_1\, .
\label{trial1.1}
\end{equation}
With the constraints in \E{twoconstraints}, the parameters $A_0$ and $B_1$ are expressed in terms of the wavenumber $l$ and substituted into \E{trial1.1}.
Thus the trial-function in \E{trial1} leads to the estimate
\begin{equation}
\R_{EZ} \approx \max_{l} \, l \sqrt{ \frac{(1 - l^2)}{2(1+l^2)} }\, .
\end{equation}
The maximum of the right hand side  is $1-2^{-1/2} =0.293 \cdots$,  which is achieved at $l^2= \sqrt{2} - 1$. This is a lower bound on $\R_{EZ}$.

To more closely approximate $\R_{EZ}$ we generalize \E{trial1} to \begin{equation}
\varphi = \cos ly \sum_{n=0}^{\infty} A_{2n} \cos 2 nx - \sin ly \sum_{n=0}^{\infty} B_{2 n+1} \sin (2 n+1) x\, .
\label{Fourier1}
\end{equation}
\T{tablez}  summarizes the result of retaining more terms in the Fourier series  \E{Fourier1}. Optimization over $A_n$, $B_n$ and $l$ provides an estimate of $\R_{EZ}$ that quickly converges to the numerical value indicated in \E{Rstarstar}. \F{EZdangerous} illustrates the most dangerous $EZ$-disturbance calculated using the final row of \T{tablez}.
\begin{figure} 
\includegraphics[width=0.44\textwidth]{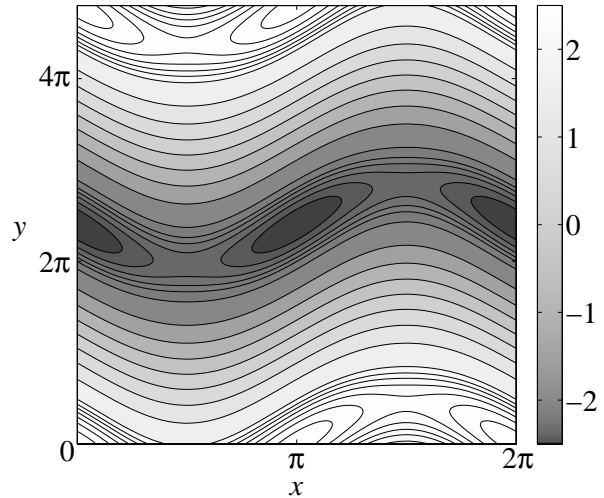}
\caption{The most dangerous $EZ$-disturbance \E{Fourier1} calculated using the final row of \T{tablez}.}
\label{EZdangerous}
\end{figure}

\begin{table}
\begin{tabular}{c c c c c c c}
$A_0$ & $A_2$ & $A_4$ & $B_1$ & $B_3$ & $l$ & $\R_{EZ}$ \\
\hline
$1.4142$ & --- & --- & $1.2872$ & --- & $0.6436$ & $0.2929$ \\
$2.1561$ & $0.2764$ & --- & $1.3010 $ & --- & $0.4253$ & 0.3556\\
$2.2141$ & $0.2694$ & --- & $1.2983$ & $0.0371$ & $0.4171$ & 0.3571 \\
$2.2175$ & $0.2689$ & $0.0037$ & $1.2984$ & $0.0376$ & $ 0.4166$ & $0.3571$\\
\hline
\end{tabular}
\caption{The estimate of $\R_{EZ}$ in the final column quickly converges to the numerical value indicated in \E{Rstarstar}
 as more terms in the Fourier series  \E{Fourier1} are retained.}
\label{tablez}
\end{table}

\section{Conclusion and  Discussion \label{conclude}}

We have studied the instability of geophysical ($\beta \neq 0$) Kolmogorov flow and focussed on the limit in which drag is much stronger than viscosity, $\mu \gg \nu$. We show that the  form of the fastest exponentially growing eigenmodes changes abruptly from stationary disturbances to traveling waves as one moves  along the neutral curve in the $(\mu,\beta)$ plane. In this sense, $\beta$ strongly affects the instability. Nonetheless, the  stability boundary obtained by the energy method \cite{Fukuta95} is reasonably close, and parallel, to the linear-stability neutral curve. Because $\beta$ enters the energy method only through the amplitude of the laminar solution i.e.,  $a(\beta, \mu,\nu)$ in \E{adef}, this success indicates that the main features of the instability are determined by $a$ and the disturbance energy equation.

 It is striking that the most unstable inviscid disturbance identified by the energy method is unphysical because it is strictly realized only in the infinite wavenumber limit, $l \to \infty$. Despite this issue, the energy method still delivers a useful sufficient condition for global monotonic stability.

In section \ref{enststab}, we extend the energy method by incorporating information based on the disturbance enstrophy equation. This EZ method results in a third stability boundary, the  EZ boundary, that lies closer to  the neutral curve than does the energy-stability boundary. One might therefore conclude that the additional information provided by enstrophy produces only a quantitative narrowing of the gap between energy stability and linear stability. However the main interest in the EZ method is that it identifies a particular  type of slowly evolving disturbances i.e., those disturbances belonging to the set $EZ$ in \E{EZset}. The EZ method relies on identifying the most unstable disturbance in $EZ$, and this disturbance, given approximately by the Fourier coefficients in \T{tablez}, has finite enstrophy.

The various types of stability we have discussed here  characterize the neighbourhood of the laminar solution $\psil(x)$ in \E{basic}. However Doering and Constantin \cite{Doering92, Doering94} have recently devised a variational procedure in which notions of energy stability are applied to statistically steady turbulent flow. This results in  bounds on important large-scale quantities, such as the mechanical energy dissipation and the heat flux. A motivation for this paper has been the possibility of applying or generalizing  the technique of Doering and Constantin to two-dimensional turbulence. In this context it is essential to take account of enstrophy conservation, perhaps via the notion of EZ-stability.

\begin{acknowledgments}

 This work was supported by the National Science Foundation by grant number OCE07-26320 and OCE02-20362.

\end{acknowledgments}

\appendix

\section{Extension of Gill's inequality to include viscosity \label{GillzAppen}}

Gill's inequality in \E{gl} restricts the Floquet wavenumber of an exponentially growing eigenmode to lie within the unit circle in the $(k,l)$-plane. To prove this result, observe that a growing eigenmode of the Floquet form \E{floquet}  satisfies \E{ddeltadt}. Thus
\begin{equation}
\left(\omega_i+ \mu \right) \left[E_{\vphi}(0) - Z_{\vphi}(0) \right] = \nu P_{\vphi}(0) - \nu Z_{\vphi}(0)\, ,
\label{gill1}
\end{equation}
where $\omega_i \geq 0$ is the imaginary part of $\omega$ and $E_{\vphi}(0)$ et cetera is the initial energy et cetera of the eigenmode. 

The disturbance enstrophy can be written as
\begin{equation}
Z_{\vphi}= - \frac12 \avg{\nabla \vphi \cdot \nabla \nabla^2 \vphi}\, ,
\end{equation} 
so that the Cauchy-Schwarz inequality implies
\begin{equation}
Z_{\vphi}^2 \leq P_{\vphi}E_{\vphi}\, .
\end{equation}
Replacing $P_{\vphi}(0)$ in \E{gill1} by $Z^2_{\vphi}(0)/E_{\vphi}(0)$ we obtain an inequality equivalent to
\begin{equation}
\left(\omega_i + \mu + \nu\frac{Z_{\vphi}(0)}{E_{\vphi}(0)} \right) \left[E_{\vphi}(0) - Z_{\vphi}(0)\right] \geq 0\, .
\end{equation}
The first factor is non-negative, and consequently a growing eigenmode must have more energy than enstrophy:
\begin{equation}
E_{\vphi} (0) \geq Z_{\vphi}(0) \, .
\label{gill2}
\end{equation}
In the inviscid case considered by Gill \cite{Gill74}, the result \E{gill2} is an equality following immediately from \E{gill1} without the  Cauchy-Schwarz excursion.

To obtain the inequality  \E{gl}, substitute the Floquet form  \E{floquet},  with
\begin{equation}
\tvphi(x) = \sum_{n=-\infty}^{\infty} \vphi_n \ee^{\ii n x}\, ,
\end{equation}
into \E{gill2}:
\begin{equation}
\sum_{n=-\infty}^{\infty} \left(1-h_n^2\right)h_n^2 |\vphi_n|^2 \geq 0\, .
\label{gill3}
\end{equation}
Above $h_n^2 \equiv (k+n)^2 + l^2$ is the wavenumber of the $n$'th wave in the Floquet series. Recalling that $-1/2 \leq k \leq 1/2$, it is easy to see that  $1-h_n^2$ is negative if $|n| \geq 2$.  Consequently  the inequality above can only be satisfied because $1-h^2_0$ or $1-h_{\pm 1}^2$ are positive. This requirement implies  \E{gl}.

\section{The inviscid energy stability eigenproblem}
\label{nu0}
In this appendix we use perturbation theory to analyze the eigenproblem \E{eigen1} in various limits. The analysis can be simplified by the transformation \cite{Fukuta95}
\begin{equation}
\tilde\vphi(x)  = \exp\left[ \frac{\ii l\lambda \sin x}{2} \right] \theta(x)\, .
\label{theta}
\end{equation}
This results in
\begin{equation}
\frac{\dd^2\theta}{\dd x^2} + 2\ii k\frac{\dd\theta }{\dd x} + \left[l^2\left(\frac{\lambda^2}{4} \cos^2x - 1\right) -k^2\right]\theta = 0 \, .
\label{thetaeq}
\end{equation}

Fukuta and Murakami \cite{Fukuta95}  solve \E{thetaeq} for the gravest eigenmode in the small wavenumber limit:
\begin{equation}
\theta = 1 + \frac14 (k^2+l^2)\cos2x  -\frac18\ii k(k^2+l^2)\sin2x 
 + O(k,l)^4\, ,
\end{equation}
with the corresponding gravest eigenvalue
\begin{equation}
\frac{\lambda_0^2}{4}\approx\left(1+\frac{k^2}{l^2}\right)\left(2-\frac{k^2+l^2}{4}\right)+ O(k,l)^4\,.
\label{lambda2}
\end{equation}
This shows that as $(k,l) \to (0,0)$,  with $l\neq 0$,  $\lambda_0 \to 2\sqrt{2}$. 

We now consider the complementary case $k=0$ and  $l\gg 1$; we define a small parameter by $\epsilon^2 \equiv l^{-1}$. Numerical solution of \E{thetaeq} indicates that the minimum of $\lambda_0(k,l)$ is at $\lambda_0(0,\infty)=2$ and we are seeking some analytic assurance of this hypothesis. The eigenfunction is concentrated in the neighbourhood of $x=0$ where $\frac{\lambda^2}{4}\cos^2x - 1$ in \E{thetaeq} is slightly positive. Thus we  expand $\cos^2 x\approx1-x^2 + O(x^4)$ and introduce a boundary-layer coordinate $X=x/\epsilon$ so that \E{thetaeq} reduces to
\begin{equation}
\theta_{XX} + \left(\frac{\lambda^2-4}{4\epsilon^2}-\frac{\lambda^2}{4} X^2\right)\theta =  O\left(\epsilon^{2}\right)\,.
\label{billzinvisc1}
\end{equation}
We recognize the quantum harmonic oscillator equation and thus obtain the eigenvalues as
\begin{equation}
\frac{\lambda_n^2-4}{2\lambda_n\epsilon^2}=2n+1\, , \qquad n=0,1,2,\dots
\end{equation}
Hence, the eigenvalue of the gravest mode ($n=0$) is
\begin{equation}
\lambda_0 \approx 2+l^{-1} + O\left(l^{-2}\right)\, ,
\label{bigl}
\end{equation}
and $\lambda_0 \to 2$ as $l\to\infty$.

The expression
\begin{equation}
\frac{\lambda_0^2}{4} = 1 + \frac{\sqrt{1 + c^2l^2}}{1+cl^2}\, , \qquad c = 1 - 2^{-1/2} =0.2928\cdots
\label{billzkluge}
\end{equation}
agrees with \E{lambda2} when $k=0$ and $l\ll 1$, and with \E{bigl} if $k=0$ and $l\gg1$. The inset of \FF{eigen_lk_nu0} shows that the interpolation \E{billzkluge} is a good approximation to the numerically computed eigenvalues for all $l$.

\section{The EZ Euler-Lagrange equation \label{EZappen}}

The direct approach to solve the EZ variational problem in section~\ref{ezvari} is to include the two constraints \E{twoconstraints} using two Lagrange multipliers $p$ and $q$. Thus setting the variational derivative of
\begin{equation}
\avg{\vphi_x \vphi_y \cos x} - q \avg{|\nabla \vphi|^2} + p\left[\avg{|\nabla \vphi|^2 - (\nabla^2 \vphi)^2}\right]
\end{equation}
to zero results in the Euler-Lagrange equation
\begin{equation}
p \nabla^4 \vphi + (p-q) \nabla^2 \vphi = \frac12\, \vphi_y \sin x -\vphi_{xy} \cos x\, .
\label{EL2}
\end{equation}
Multiplying \E{EL2} by $\vphi$ and taking the spatial average, we deduce that
\begin{equation}
q = \avg{\vphi_x \vphi_y \cos x} \, .
\end{equation}
Writing $\vphi$ in the Floquet form \E{floquet} (with $\omega=0$), we solve \E{EL2} by regarding $p$ as an eigenvalue and $q$ as a parameter, for different values of $k$ and $l$ satisfying the Gill's inequality \E{gl}. For a given $q$, admissible solutions are the subset of eigenfunctions of \E{EL2} that satisfies
\begin{equation}
\frac{\avg{|\nabla\vphi|^2}}{\avg{(\nabla^2 \vphi)^2}} - 1 = 0\,.
\label{cq}
\end{equation}
$\R_{EZ}$, defined in \E{Rstarstar}, is then given by the maximum $q$ at which such solutions exist.
Now, we know that
\begin{equation}
0.3571 \leq q \leq 0.5\, .
\label{EL4}
\end{equation}
The upper limit is what we get if we ditch the enstrophy constraint and perform the maximization within the larger class of functions satisfying only $ \avg{ | \nabla \vphi|^2 } =1$. The lower limit is obtained from the trial function method described in section~\ref{ezvari}. Thus, we only need to search for $\R_{EZ}$ within the range in \E{EL4}. The result is $\R_{EZ}=0.3571$ with $p=0.0069$, this occurs at $k=0$ and $l=0.4166$. These numbers are virtually the same as the trial function result given in the final row of \T{tablez}. We observe that in both the energy method and the EZ method, the most unstable disturbance has wavenumber $k=0$, we have not been able to prove this result analytically.

\bibliography{kolstab5}
\bibliographystyle{apsper.bst}

\end{document}